\documentclass[twocolumn,aps,prl,superscriptaddress]{revtex4-2}

\usepackage{amsmath}
\usepackage{graphicx}
\usepackage{MnSymbol}
\usepackage{parskip}
\usepackage{makecell}
\usepackage{booktabs}
\usepackage{multirow}

\begin{document}

\title{Multiscale contact mechanics for elastoplastic contacts}

\author{A. Almqvist}
\affiliation{Division of Machine Elements, Lule{\aa} University of Technology, 97187, Lule{\aa}, Sweden}

\author{B.N.J. Persson}
\affiliation{State Key Laboratory of Solid Lubrication, Lanzhou Institute of Chemical Physics, Chinese Academy of Sciences, 730000 Lanzhou, China}
\affiliation{Peter Gr\"unberg Institute (PGI-1), Forschungszentrum J\"ulich, 52425, J\"ulich, Germany}
\affiliation{MultiscaleConsulting, Wolfshovener str. 2, 52428 J\"ulich, Germany}

\begin{abstract}
Understanding contact between rough surfaces undergoing plastic deformation is crucial in many applications. We test Persson’s multiscale contact mechanics theory for elastoplastic solids, assuming a constant penetration hardness. Using a numerical model based on the boundary element method, we simulate the contact between a flat rigid surface and an elastic-perfectly plastic half-space with a randomly rough surface. The theory's predictions for elastic, plastic, and total contact area agree quantitatively with the numerical results. The simulations also support the boundary conditions assumed in the theory, namely that the stress probability vanishes at both zero and yield stress. These findings reinforce the validity of the theory for systems with constant hardness.
\end{abstract}

\maketitle

\setcounter{page}{1}
\pagenumbering{arabic}

%\pagestyle{empty}

%%%%%%%%%%%%%% main text %%%%%%%%%%%%%%%%
%\begin{multicols}{2}

%%%%%%%%%%%%%% main text %%%%%%%%%%%%%%%%

\section{Introduction} 

All solid bodies have surface roughness. When two solids are squeezed together, they will usually only make contact in a small fraction of the nominal contact region. In many cases, the local stresses in the asperity contact regions are so high that the solids yield plastically, at least in some fraction of the asperity contact regions. Understanding how to determine the contact area and the distribution of contact stress for elastoplastic contact is an important topic for a large number of practical applications.

One analytical approach to the elastoplastic contact between solids is the Persson contact mechanics theory. This theory in its simplest form assumes that the solids yield plastically when the local stress reaches the penetration hardness $\sigma_{\rm P}$. The penetration hardness is determined in indentation experiments, where a very hard object (often diamond) of well-defined shape (usually a ball or pyramid) is pressed into contact with the solid. The ratio between the loading force and the area of the (plastically deformed) indentation is defined as $\sigma_{\rm P}$. The elastoplastic contact mechanics theory has been applied to the leakage of metallic seals~\cite{Metal} and to the leakage of syringes with Teflon laminated rubber stoppers~\cite{Nest}. In a recent study, Lambert and Brodsky~\cite{Brodsky} have applied the theory (with a length-scale dependent penetration hardness) to the contact between the surfaces in earthquake faults.

Consider the contact between two elastoplastic solid bodies with nominal flat surfaces with roughness.
When the interface is observed at the magnification $\zeta$ only the roughness components with wavelength  $\lambda > L/ \zeta$, can be observed,
where where $L$ is the linear size of the surface. We define the wavenumber $q=2\pi/\lambda$ and the reference wavenumber $q_0 = 2\pi/L$.

The Persson contact mechanics theory is based on studying the probability distribution $P(\sigma,\zeta)$ of contact stress $\sigma$ as the magnification $\zeta$ increases. This function obeys a diffusion-like equation where time is replaced by magnification $\zeta$ and the spatial coordinate by the stress $\sigma$, and where the diffusivity depends on the surface roughness power spectrum and on the elastic properties of the solids. Solving this equation requires boundary conditions. Since there can be no negative stress at the interface (assuming no adhesion) $P(\sigma, \zeta) = 0$ for $\sigma < 0$. In fact, one can show that for elastic contact $P(\sigma, \zeta) \rightarrow 0$ as $\sigma \rightarrow 0^+$. For elastoplastic contact, the same condition is assumed to hold, with the additional assumption that $P(\sigma, \zeta) \rightarrow 0$ as $\sigma \rightarrow \sigma_{\rm P}^-$, where $\sigma_{\rm P}$ is the penetration hardness.
Here we will show, by comparing the analytical theory with exact numerical simulation results, that both boundary conditions hold for elastoplastic contact.

The original derivation of the stress probability distribution in the asperity contact regions for elastoplastic contact resulted in an 
infinite series~\cite{JCPP}. In a recent study, Xu et al.~\cite{ep1} have derived a simpler analytical expression for the stress probability distribution. 
%This study also claims to have proved the correctness of the boundary conditions $P(\sigma, \zeta) \rightarrow 0$ as $\sigma \rightarrow 0^+$ and $\sigma \rightarrow \sigma_{\rm P}^-$. However, the proof is not clear to us, as it appears to make use of the solution obtained, assuming the correctness of the boundary conditions.

Venugopalan et al.~\cite{ep2} compared the Persson elastoplastic theory predictions with a numerical calculation using a discrete dislocation plasticity model. This model incorporates strain hardening, implying that the effective penetration hardness depends on the indentation depth, or equivalently, on the length scale or magnification. As a result, no detailed comparison with Persson's theory was possible, but the numerical results supported the assumption that $P(\sigma, \zeta) \rightarrow 0$ as $\sigma \rightarrow 0^+$. We note that the Persson theory can be extended to include a dependency of the penetration hardness on the magnification~\cite{ep5, Brodsky}, but this more complete theory approach was not used in Ref.~\cite{ep2},
probably because it is much harder to implement numerically. It also requires as input the dependency of the penetration hardness on the magnification, but this could have been obtained in Ref.~\cite{ep2} by performing (numerical) indentation simulations with a rigid indenter.

The aim of this study is to perform a rigorous test of the Persson elastoplastic theory in the simplest case of a constant penetration hardness. We compare the theory predictions to exact boundary element calculations, which are possible for systems with roughness over a small length scale region. We show that the theory predicts accurately both the elastic and plastic contact area, and also support the assumption that $P(0,\zeta) = P(\sigma_{\rm P}, \zeta)=0$ are the correct boundary conditions to use when solving the stress probability diffusion equation.

\begin{figure}
\includegraphics[width=0.25\textwidth,angle=0.0]{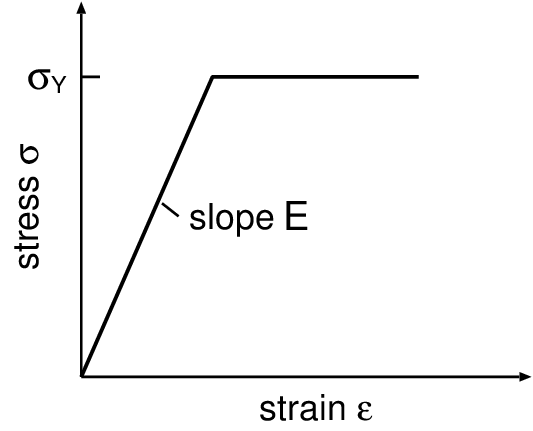}
\caption{\label{ElastoPlasticStressStrain.eps}
The relation between stress and strain in uniaxial tension for the simplest elastoplastic material model. The material deforms elastically with slope $E$ up to the yield stress $\sigma_{\rm Y}$, beyond which it flows plastically without strain hardening.
}
\end{figure}

\vskip 0.3cm
\section{Multiscale contact mechanics theory} 

We study the interface between an elastoplastic solid with a randomly rough surface \cite{almqvist_persson_2026_8192_32_digital_surface}, and a rigid body with a perfectly flat surface at a magnification $\zeta$. At this magnification, we only observe the roughness with wavenumber components below $\zeta q_0$, where $q_0$ is the wavenumber of the smallest roughness component. We assume that the solids have nominally flat surfaces and are squeezed together with a nominally uniform stress. The nominal contact area is denoted by $A_0$. Because of the surface roughness, the stress acting at the interface is highly non-uniform. In the Persson contact mechanics theory for elastic solids, the probability distribution of contact stress equals:
\begin{equation}\label{eq1}
    P(\sigma,\zeta) = \dfrac{1}{(4\pi G)^{1/2}} \left (e^{-(\sigma-\sigma_0)^2/4G}- e^{-(\sigma+\sigma_0)^2/4G} \right ),
\end{equation}
where $\sigma_0$ is the nominal (applied) pressure, and where
\begin{equation}\label{eq2}
    G (\zeta) =\frac{\pi}{4} (E^*)^2 \int_{q_0}^{\zeta q_0} dq \ q^3 C(q)S(q),
\end{equation}
where $E^* = E/(1-\nu^2)$ is an effective elastic modulus and $C(q)$ the surface roughness power spectrum.

The factor $S(q)$ in (2) takes into account that the elastic energy stored in the contact regions is smaller than what would be the case if complete contact would occur
in the contact regions observed at the magnification $\zeta$.
When the relative contact area $A/A_0 << 1$, as in the applications below, $S(q)\approx\gamma$ (see Ref.~\cite{ep11}). 
The elastic energy reduction factor $\gamma$ has been determined originally by comparing the average surface separation as a function of squeezing pressure with numerical simulation results, and has been found to be (see Ref.~\cite{ep11}) $\gamma\approx 0.5$. Here we will use $\gamma=0.536$. Note that using the correction factor $S(q)\approx\gamma$ is equivalent to using $S=1$ and a reduced modulus $E_{\rm eff} \approx E \surd \gamma \approx 0.73 E$.

When the stress in the asperity contact region becomes high enough, plastic flow occurs. In the simplest model, it is assumed that a material deforms as a linear elastic solid until the stress reaches a critical level, the so-called plastic yield stress, where it flows without strain hardening, see Fig.~\ref{ElastoPlasticStressStrain.eps}. The yield stress in elongation is denoted by $\sigma_{\rm Y}$. In indentation experiments, where a sharp tip or a sphere is pushed against a flat solid surface, the penetration hardness $\sigma_{\rm P}$ is defined as the ratio between the normal force and the projected (on the surface plane) area of the plastically deformed indentation. As shown by Tabor\cite{Tabor}, typically $\sigma_{\rm P} \approx 3 \sigma_{\rm Y}$. We note that the yield stress of materials often depends on the length scale (or magnification), which in principle can be included in the formalism we use \cite{Preview,Brodsky}.

The influence of plastic flow on the contact mechanics is taken into account in the Persson contact mechanics approach by replacing the boundary condition $P(\infty,\zeta) = 0$ with $P(\sigma_{\rm Y},\zeta) = 0$. Note that although there can be no stress larger than the penetration hardness, i.e., $P(\sigma, \zeta) = 0$ for $\sigma > \sigma_{\rm P}$, this does not by itself guarantee that $P(\sigma,\zeta)\rightarrow 0$ as $\sigma \rightarrow \sigma_{\rm P}^-$. The same is true as $\sigma \rightarrow 0^-$ so the boundary conditions $P(\sigma_{\rm Y},\zeta) = 0$ and $P(0,\zeta) = 0$ are non-trivial. This approach is based on the simplest elastoplastic description, where only elastic deformation occurs for $\sigma < \sigma_{\rm P}$, while for $\sigma = \sigma_{\rm P}$, the material flows without work-hardening so that the maximal stress equals $\sigma_{\rm P}$ (see Fig.~\ref{ElastoPlasticStressStrain.eps}).

The pressure probability distribution in the elastoplastic theory, in the region where elastic deformation has occurred, is given by~\cite{JCPP}:
\begin{equation}\label{eq3}
    P(\sigma,\zeta) = \frac{2}{\sigma_{\rm P}} \sum_{n=1}^\infty \, {\rm sin} (s_n\sigma_0) \, {\rm sin} (s_n\sigma) \, e^{-s_n^2 G(\zeta)},
\end{equation}
where $s_n = n \pi/\sigma_{\rm P}$. 
Note that, as $\sigma_{\rm P} \rightarrow \infty$, \eqref{eq3} reduces to \eqref{eq1}. 
If $A_{\rm el} (\zeta) /A_0$ is the fraction of the nominal area where the solids are in elastic contact when the interface is observed at the magnification $\zeta$, then 
\begin{align}
    &\frac{A_{\rm el} (\zeta )}{A_0} = \int_0^{\sigma_{\rm P}} d \sigma \ P(\sigma,\zeta) \nonumber \\[2ex]
    & = \frac{2}{\pi} \sum_{n=1}^\infty \, \frac{1}{n} \left[ 1-(-1)^n\right ] {\rm sin} (s_n\sigma_0) \, e^{-s_n^2 G(\zeta)} \label{eq4}
\end{align}
The fraction of the nominal area where the solids are in plastic contact when the interface is observed at the magnification $\zeta$ is
\begin{equation}\label{eq5}
    \frac{A_{\rm pl} (\zeta )}{A_0} = \frac{\sigma_0}{\sigma_{\rm P}}+\frac{2}{\pi} \sum_{n=1}^\infty \, \frac{(-1)^n}{n} {\rm sin} (s_n\sigma_0) \, e^{-s_n^2 G(\zeta)}.
\end{equation}

\vskip 0.3cm
\section{Boundary element method} 

The numerical model employed in this work simulates two-dimensional frictionless contact between an elastic-perfectly plastic half-space with a randomly rough surface and a rigid body with a flat surface, using a spectral boundary element method grounded in the framework introduced by Almqvist et al.~\cite{Almqvist2007}. Surface roughness is represented as a periodic height field, and deformations are computed in the frequency space,  via convolution with the Green's function of an elastic half-space. Plastic deformation is introduced through a local yield pressure: wherever the contact pressure exceeds this threshold, the node is treated as plastically loaded and constrained to maintain contact without bearing further load. The solver employs relaxation-based iteration with residual convergence to enforce force balance and has been furthered and applied in several other works, e.g., \cite{Sahlin2010, Almqvist2011, metal1, PerezRafols2021, Kalliorinne2021, Kalliorinne2022, kalliorinne2023}.

\begin{figure}
\includegraphics[width=0.45\textwidth,angle=0.0]{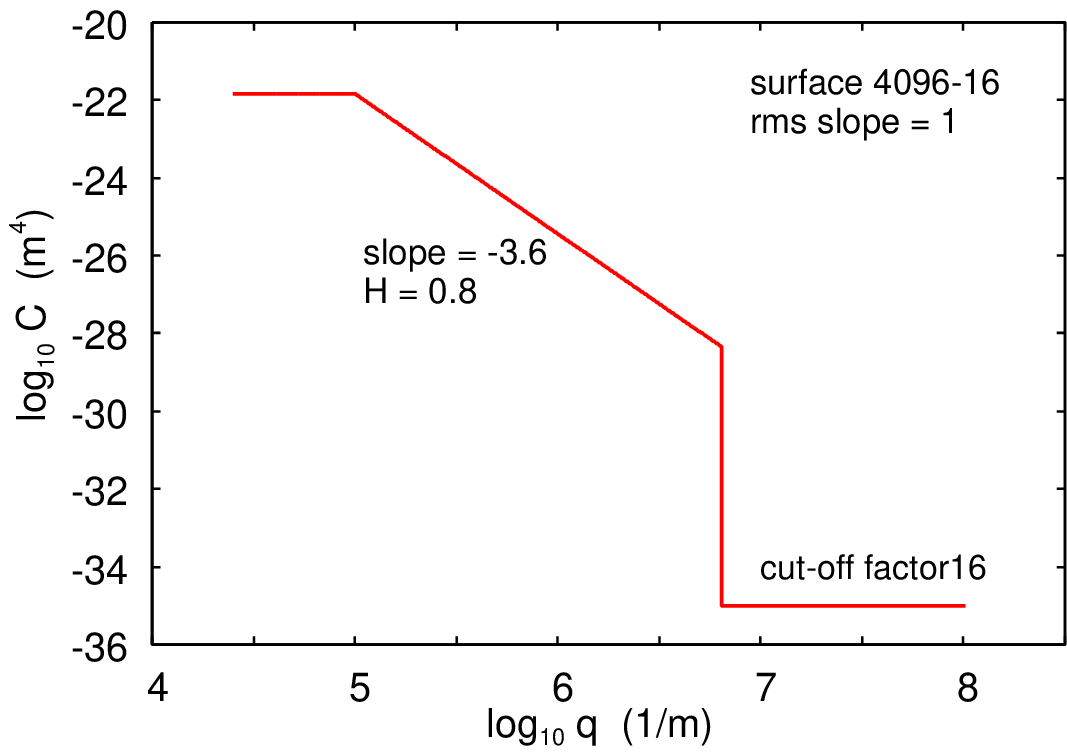}
\caption{\label{1logq.2logC.4096.eps}
The surface roughness power spectrum as a function of the wavenumber (log-log scale) for the surface 4096-16. The surface has rms-slope $\xi =1$, and the ratio between the highest and smallest wavenumber is $q_1/q_0 = 4096$. The ``width'' of the large wavenumber cut-off region is $16$.
}
\end{figure}

\begin{figure}
\includegraphics[width=0.45\textwidth,angle=0.0]{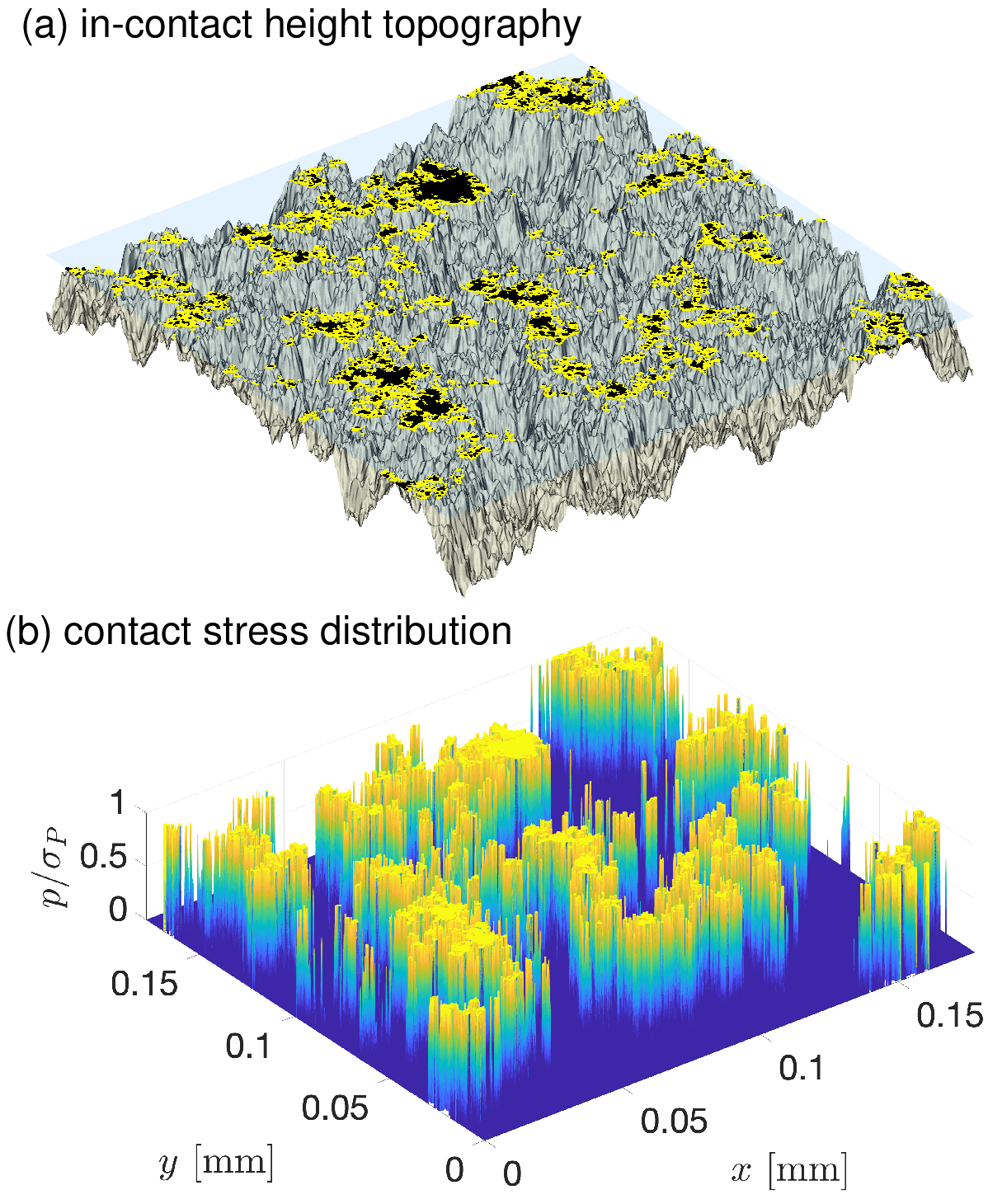}
\caption{\label{TopographyStressPic}
The calculated (a) in-contact topography squeezed against a rigid flat surface (transparent), and (b) contact stress distributions. The results are obtained using the boundary element method (BEM) for an elastoplastic model with Young's modulus $E=250\ \mathrm{GPa}$, Poisson ratio $\nu=0$, and the yield stress (or penetration hardness) $\sigma_{\rm P} = 150\ \mathrm{GPa}$. The applied squeezing pressure is $\sigma_0 = 12.50\ \mathrm{GPa}$. The calculated relative elastic contact area is $A_{\rm el}/A_0 = 0.0533$ (yellow), plastic contact area $A_{\rm pl}/A_0 = 0.0572$ (black), and total contact area $A/A_0 = 0.1105$.}
\end{figure}

\begin{figure}
\includegraphics[width=0.45\textwidth,angle=0.0]{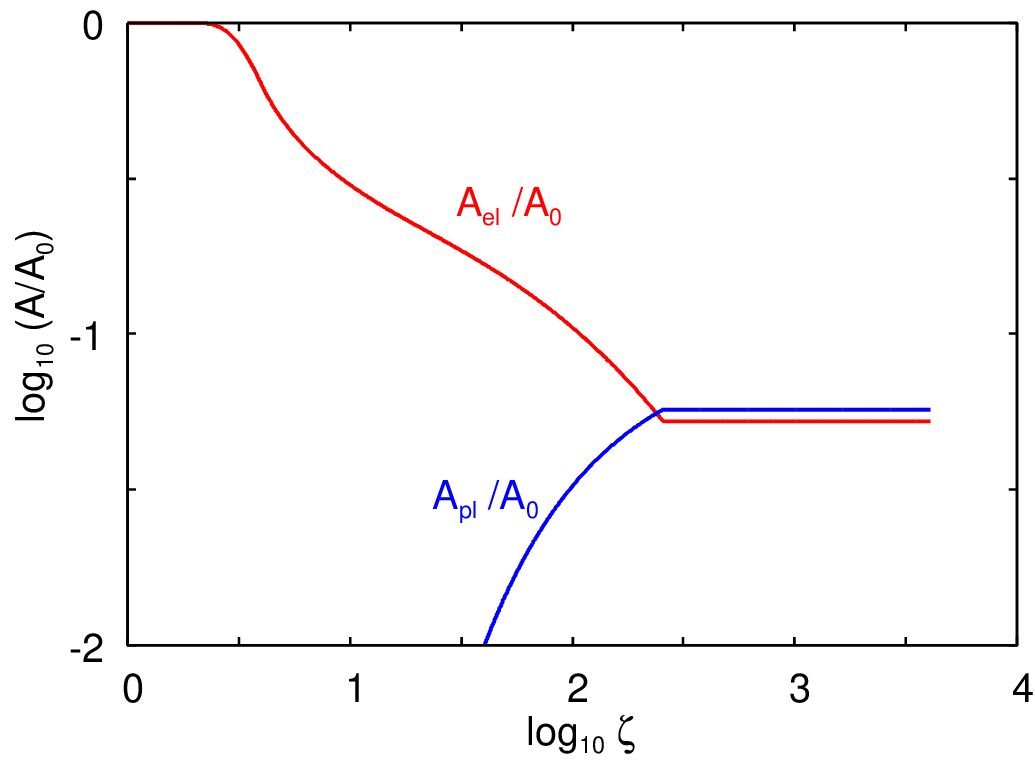}
\caption{\label{1logZETA.2logArea.Ael.0.524.Apl.0.573.eps}
The elastic and plastic contact area as a function of the magnification (log-log scale), as predicted by the analytical theory. The true contact areas are obtained at the highest magnification and are $A_{\rm pl}/A_0 = 0.0572$ and $A_{\rm el}/A_0 = 0.0526$. 
The total contact area $A/A_0 = (A_{\rm el} + A_{\rm pl})/A_0 =0.1099$. For the same applied pressure and the same material parameters as in Fig.~\ref{TopographyStressPic}. The elastic energy reduction factor $\gamma = 0.536$.}
\end{figure}

\begin{figure}
\includegraphics[width=0.45\textwidth,angle=0.0]{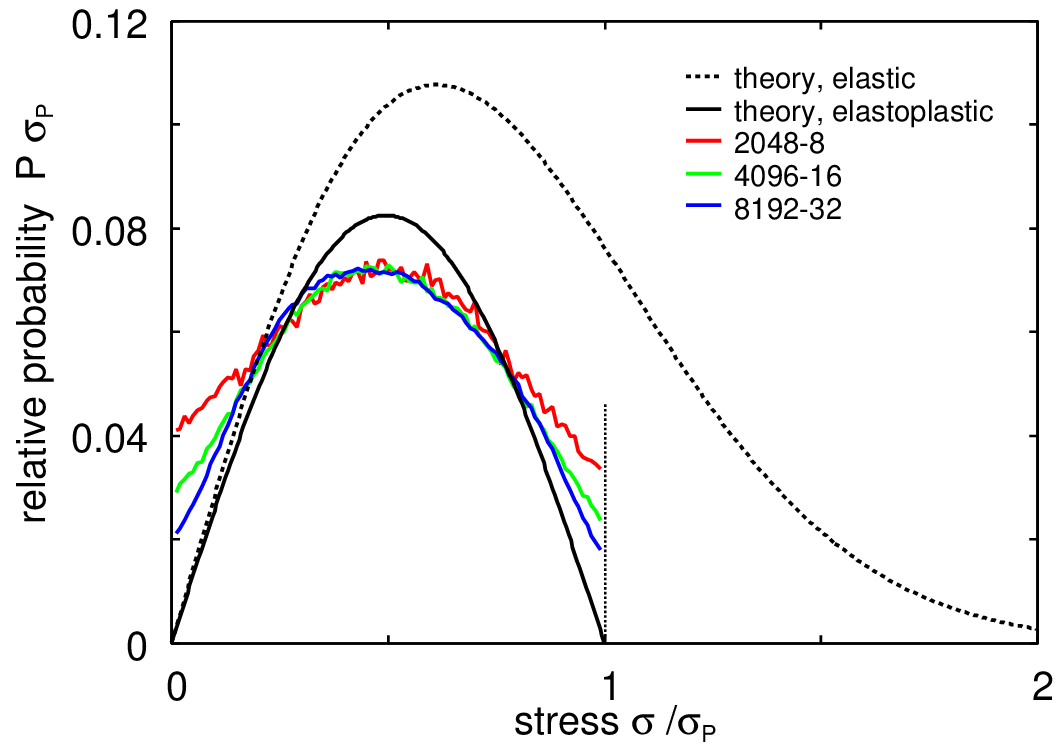}
\caption{\label{1pressure.1Probability.improved1.eps}
The black dotted line is the probability distribution of contact stresses for elastic contact with the same parameters as in  Fig.~\ref{TopographyStressPic} 
except with $\sigma_{\rm P} = \infty$. The block solid line is the probability distribution of elastic stress from the analytical theory. The red, green and blue lines are the probability distribution of elastic stress from the boundary element calculations for the 2048-8, 4096-16 and 8192-32 systems, respectively. Not shown in the figure are the Dirac delta functions $\sim \delta(\sigma )$ and $\sim \delta (\sigma-\sigma_{\rm P})$, corresponding to the non-contact area and plastically deformed contact area, respectively. In the calculations of the solid lines, we have used the same applied pressure and the same material parameters as in Fig.~\ref{TopographyStressPic}.}
\end{figure}

\vskip 0.3cm
\section{Comparison of theory with numerical results} 

We have studied the contact mechanics of an elastoplastic solid with a randomly rough surface being squeezed against a rigid substrate with flat surface. The rough surfaces have self-affine fractal roughness in a wavenumber interval $1 \times 10^5 \ {\rm m}^{-1} < q < 64 \times 10^5 \ {\rm m}^{-1}$, with the fractal dimension $D_{\rm f}=2.2$ (corresponding to the Hurst exponent $H=0.8$). We have added a short roll-off region for $2.5 \times 10^4 \ {\rm m}^{-1} < q < 1 \times 10^5 \ {\rm m}^{-1}$ as this generate some self averaging, and in addition, most surfaces of engineering interest have a roll-off region in the power spectra (since most surfaces are designed to be smooth at the macroscopic length scale)~\cite{RollOff}. When comparing numerical simulations with continuum mechanics theories, it is crucial to have enough grid points within the shortest roughness wavelength. This can be obtained by using surfaces which are generated from power spectra that have a large enough cut-off region for large wavenumber, as illustrated in Fig.~\ref{1logq.2logC.4096.eps} for a surface where the ``width'' of the cut-off region is $16$. The total ``length'' of the power spectrum is $q_1/q_0 = 4096$, so we denote the surface as 4096-16. The power spectrum shown in Fig.~\ref{1logq.2logC.4096.eps} has the rms-slope $\xi =1$. We consider surfaces with power spectra identical to the one shown in Fig.~\ref{1logq.2logC.4096.eps}, except for the differing lengths of the large wavenumber cut-off regions. All these surfaces have the rms-slope $\xi\approx1$ since the cut-off region contributes negligibly to the rms slope.

In all calculations we use the Young's modulus $E=250 \ {\rm GPa}$, the Poisson ratio $\nu=0$ and the penetration hardness $\sigma_{\rm P} = 150 \ {\rm GPa}$ (corresponding to a Yield stress in tension $\approx 50 \ {\rm GPa}$). In elastic contact mechanics, as long as the contact area $A<<A_0$, the contact area is proportional to the nominal squeezing pressure $\sigma_0$:
\begin{equation}\label{eq6}
    \frac{A}{A_0} \approx \frac{2 \sigma_0}{\xi E^*},
\end{equation}
where $E^*=E/(1-\nu^2)$. In the present case, $E^*=E= 250 \ {\rm GPa}$ and $\xi=1$. Hence, if the applied stress $\sigma_0 = 12.5 \ {\rm GPa}$, then $A/A_0 = 0.1$ for elastic contact, and this is the applied stress we will use in what follows.

We have generated randomly rough surfaces $z=h(x,y)$ with the power spectra given above by adding plane waves with random phases (see Appendix A in Ref.~\cite{Tosatti}):
\begin{equation}\label{eq7}
    h({\bf x}) = \sum_{\rm q} B({\bf q}) e^{i[{\bf q}\cdot {\bf x}+\phi_{\bf q}]}
\end{equation}
where $\phi_{\bf q}$ are random numbers uniformly distributed between $0$ and $2\pi$.
In \eqref{eq7}
\begin{equation}\label{eq8}
    B({\bf q}) = \frac{2\pi}{L} [C({\bf q})]^{1/2},
\end{equation}
where $L^2=A_0$ is the nominal surface area. The wavevector ${\bf q}$ spans all the vectors whose components are whole multiples of $2\pi/L$ but with magnitude smaller than the cut-off wavenumber $q_1$.

Fig.~\ref{TopographyStressPic}(a) shows the calculated height topography when the elastic block is squeezed against a surface with the power spectrum 4096-16. The results are obtained using the boundary element method (BEM) for an elastoplastic model with the parameters given above. Fig.~\ref{TopographyStressPic}(b) shows the stress acting in the area of real contact. From the stress distribution, for the 8192-32 system one can obtain the relative elastic contact area $A_{\rm el}/A_0 = 0.0533$, and the plastic contact area $A_{\rm pl}/A_0 = 0.0572$. The total contact area $A/A_0 = (A_{\rm el} + A_{\rm pl})/A_0 = 0.1105$. We compare these values with the theoretical predictions.

Fig.~\ref{1logZETA.2logArea.Ael.0.524.Apl.0.573.eps} shows the elastic and plastic contact area as a function of the magnification as predicted by the analytical theory \eqref{eq4} and \eqref{eq5}. The true contact areas are obtained at the highest magnification and are $A_{\rm el}/A_0 = 0.0526$ and $A_{\rm pl}/A_0 = 0.0572$. The total contact area $A/A_0 = (A_{\rm el} + A_{\rm pl})/A_0 =0.1099$. These numbers are in nearly perfect agreement with the numerical simulation results, see Table~\ref{tab1}. In the table we also show the relative contact area $A/A_0$ for elastic contact, i.e., when $\sigma_{\rm Y} = \infty$, but with all the other material parameters the same as for the elastoplastic contact. The elastic contact mechanics case has been studied in great detail elsewhere~\cite{Martin1,Martin2} and will not be discussed further here.

\setlength{\tabcolsep}{6pt}
\begin{table}[h!]
    \centering
    \vspace{5pt}
    \begin{tabular}{lccccc@{\hskip 10pt}c}
    \toprule
    & & \multicolumn{3}{c}{elastoplastic} & & elastic \\
    \cmidrule(lr){3-5} \cmidrule(lr){7-7}
    & \makecell{Cut-off\\length} & $A_{\mathrm{el}}/A_0$ & $A_{\mathrm{pl}}/A_0$ & $A/A_0$ & & $A/A_0$ \\
    \midrule
    \multirow{3}{*}{BEM}
    & 8   & 0.0576 & 0.0551 & 0.1126 & & 0.1028 \\
    & 16 & 0.0542 & 0.0568 & 0.1109 & & 0.1003 \\
    & 32 & 0.0533 & 0.0572 & 0.1105 & & 0.1002 \\
    \midrule
    Theory & $\infty$ & 0.0526 & 0.0572 & 0.1099 & & 0.1087 \\
    \bottomrule
    \end{tabular}
    \caption{Comparison of fractional elastic ($A_{\mathrm{el}}/A_0$), plastic ($A_{\mathrm{pl}}/A_0$), and total contact area ($A/A_0$) for
    elastoplastic contact for surfaces with different cut-off lengths. The last column shows the same for elastic contact.}
    \label{tab1}
\end{table}

Fig.~\ref{1logZETA.2logArea.Ael.0.524.Apl.0.573.eps} shows that at low magnification, where only the long-wavelength roughness can be observed, it appears as if the contact is purely elastic. Only at high enough magnification can one observe a plastically deformed contact area. Thus, the theory predicts the length scale at which plastic flow becomes important, providing very useful information in many applications. Note also that when the magnification becomes so large that the $\zeta q_0$ enter the cut-off region in Fig.~\ref{1logq.2logC.4096.eps} the contact area does not change any more as the magnification increases. This is due to the (nearly) vanishing amplitude of these short wavelength roughness components.

Next, let us compare the stress probability distributions calculated based on the results obtained with the BEM-based model with the theoretical predictions. The dotted black line in Fig.~\ref{1pressure.1Probability.improved1.eps} is the probability distribution of contact stresses for elastic contact with the same parameters as in  Fig.~\ref{TopographyStressPic} except with $\sigma_{\rm P} = \infty$. The black solid line is the probability distribution of elastoplastic contact stress predicted with the analytical theory. The green, red, and blue lines are the probability distribution of elastoplastic contact stress obtained with the BEM-based model for the 2048-8, 4096-16,  and 8192-32 systems, respectively. With the present numerical method, the way it is implemented, we are not able to perform studies of systems with a larger cut-off region, but the figure makes it plausible that as the cut-off region increases, the stress probability distribution approaches the continuum prediction, and in particular 
$P(\sigma, \zeta) \rightarrow 0$ as $\sigma \rightarrow 0^+$ and $\sigma \rightarrow \sigma_{\rm P}^-$.

\vskip 0.3cm
\section{Discussion}

We have compared Persson's contact mechanics theory with a numerical BEM-based method that incorporates a simple model of plasticity, where solids deform as linear elastic materials until the stress reaches the penetration hardness, after which plastic flow occurs without work (or strain) hardening.  

A key result of our study is that we demonstrate how the boundary conditions $P(\sigma, \zeta) \to 0$ as $\sigma \to 0^+$ and $\sigma \to \sigma_{\rm P}^-$, which are assumed in the analytical theory, naturally emerge from the boundary element simulations. These conditions are not imposed explicitly; rather, they follow from the way plasticity is implemented. Specifically, nodes where $\sigma > \sigma_{\rm P}$ are classified as plastically loaded and remain in contact without bearing additional load, while nodes with $0 < \sigma < \sigma_{\rm P}$ and $\sigma = 0$ are considered elastically loaded and non-contacting, respectively. While this setup constrains the stress distribution to lie within the physically admissible range, it does not, \textit{a priori}, enforce that $P(\sigma, \zeta)$ vanishes continuously at the boundaries. Nevertheless, the numerical results shown in Fig.~\ref{1pressure.1Probability.improved1.eps} clearly exhibit this behavior, thereby validating the theoretical boundary conditions in the continuum limit. 
%This contrasts with the work of Xu et al.~\cite{ep1}, who assumed these boundary conditions without providing a rigorous justification in the elastoplastic case.

Xu et~al.~\cite{ep1} derived a simple closed-form expression for the stress probability distribution,  $P(\sigma,\zeta)$, that satisfies these boundary conditions. They also presented a consistency argument in support of them. However, because their closed-form solution was itself constructed by imposing these conditions, the resulting argument does not constitute an independent derivation of the boundary conditions $P(\sigma,\zeta)\to 0$ as $\sigma\to 0^{+}$ and $\sigma\to\sigma_{\rm P}^{-}$ in Persson's elastoplastic contact mechanics theory. Our BEM results demonstrate these limits naturally without such an assumption.

An important parameter in the elastoplastic theory is the penetration hardness and here will discuss it in some detail. Indentation hardness can be measured using indenters of various sizes and shapes. Here we define the penetration hardness as the applied force divided by the projected indentation area. In general, the penetration hardness depends on the shape of the indenter (because of the different form of the stress field and plastic deformation)~\cite{review}. Macroscopic tests usually use spherical or pyramidal shapes. Most rough surfaces, when observed at any given magnification, have ``smooth'' asperities like spherical cups (sharp spike-like structures are unlikely as they would be removed during the surface preparation, such as polishing), and we believe using a spherical indenter (as in the Brinell hardness test) gives a penetration hardness most suitable for contact mechanics calculations.

The penetration hardness often depends on the size of the indenter and on the indentation depth. In nanoindentation studies, it has also been demonstrated that hardness depends on the indentation depth, particularly in the \emph{nanometre} range. To establish a useful correlation of hardness with material flow properties, it is essential to characterize the variation in hardness with indentation depth. An increase in hardness with decreasing indentation depth has been observed in numerous nanoindentation experiments on various materials~\cite{indent0, indent1, indent2, indent3}. This increase is associated with the mechanism of plastic flow, which in crystalline solids involves the creation and motion of dislocations.

The contact mechanics theory developed in Refs.\cite{ep4, Preview} allows for a penetration hardness which varies with the length scale. The implementation of this in a computer code is cumbersome, but was done in a recent paper by Lambert and Brodsky\cite{Brodsky}, who applied the theory in a study of how plastic yielding of natural surfaces, particularly earthquake fault surfaces, may modify the surface topography at different length scales. Brodsky et al.\cite{Brod} have proposed the intriguing question of whether scale-dependent strength can be inferred from the study of the topography of natural surfaces. 

The importance of scale (or magnification) dependent hardness was observed in a recent study by Patil et al.~\cite{Ali}. They studied the dependency of the contact area and the interfacial separation on the normal force when a germanium spherical crystal was squeezed against a rough glassy polymer (PMMA) surface. They found that the dependency of the contact area on the normal force cannot be explained by assuming a constant penetration hardness and suggested that a depth-dependent hardness must be used, as also suggested in a study by Weber et al.~\cite{Weber}.

The importance of work (or strain) hardening (which in crystalline solids is due to the intersection of dislocations on different slip planes) in contact mechanics was shown in a recent study by Venugopalan et al.~\cite{ep2}. They compared the Persson contact mechanics prediction with a constant penetration hardness to numerical simulations using a discrete dislocation plasticity model where strain hardening occurs. It would have been interesting to extend this study by using a scale-dependent penetration hardness, which could be obtained from the numerical study by performing indentations of varying depths.

Even without strain hardening the effective penetration hardness may depend on the amount of plastic deformation. 
Thus, it has been known for a long time that the asperities do not flatten out perfectly, but only the upper part of some fraction of the asperities flatten 
in such a way that the flattened area typically occupies half of the macroscopic indentation area \cite{A23,A24,A25}. 
This persistence of asperities in indentation experiments can be interpreted as resulting from an indentation hardness, which increases as the length scale (or indentation size) decreases, but a more likely explanation is related to plasticity mechanics of asperity interaction as proposed in \cite{A23,A24,A28,A29,A30}. That is, qualitatively, as an asperity becomes strongly plastically deformed, the stress field approaches a hydrostatic stress, and the asperity therefore becomes resistant to further plastic deformation. This is not a length scale effect and cannot be simply included as a dependency of the penetration hardness on the length scale of the roughness or magnification.

The penetration hardness depends on so many factors that, in most cases, it cannot be predicted theoretically, but needs to be measured experimentally. In spite of all the ``complications'' discussed above, we believe that contact mechanics calculations using the model reported on in Sec.~2, possibly with a length scale (or magnification) dependent penetration hardness, may be extremely useful for understanding physical phenomena such as heat and electric contact resistance, which depend on the area of real contact and on the size of the contact regions.

\vskip 0.3cm
\section{Summary and conclusion} 

We have tested Persson's multiscale contact mechanics theory for elastoplastic solids, assuming a constant penetration hardness. Using a spectral boundary element method, we simulated contact between a elastic–perfectly plastic half-space with randomly rough surfaces, and a rigid substrate with a flat surface. The theory's predictions for elastic, plastic, and total contact areas are in excellent quantitative agreement with the numerical results. Moreover, the simulations confirm that the boundary conditions assumed in the theory, i.e., that the stress probability distribution vanishes at both zero and yield stress, arise naturally in the numerical model. These findings validate the theoretical framework for systems with constant hardness and support its use in practical contact mechanics applications. While scale-dependent hardness may be important in some cases, the constant-hardness model captures the dominant behavior in a wide range of scenarios.

%\bibliographystyle{apsrev4-2}
%\bibliography{references}  % references.bib file (without .bib extension)

\end{document}